\documentclass[aps,prd,a4paper,nosuperscriptaddress,nofootinbib,showpacs,twocolumn,showkeys,amsfonts,amssymb,amsmath]{revtex4-1}
\usepackage{amssymb,latexsym}
\usepackage{amsmath, amsthm}

\usepackage{color}
\usepackage{amscd}
\usepackage{times}
\usepackage{epsfig}
\usepackage{psfrag}
\usepackage{graphicx}
\newcommand{\bea}{\begin{eqnarray}}
\newcommand{\eea}{\end{eqnarray}}
\newcommand{\be}{\begin{equation}}
\newcommand{\ee}{\end{equation}}

\begin{document}
\title[]{Singularity avoidance in quantum-inspired inhomogeneous dust collapse}

\author{Yue Liu}
\affiliation{Center for Field Theory and Particle Physics \& Department of Physics,
Fudan University, 220 Handan Road, 200433 Shanghai, China}

\author{Daniele Malafarina}
\affiliation{Center for Field Theory and Particle Physics \& Department of Physics,
Fudan University, 220 Handan Road, 200433 Shanghai, China}

\author{Leonardo Modesto}
\affiliation{Center for Field Theory and Particle Physics \& Department of Physics,
Fudan University, 220 Handan Road, 200433 Shanghai, China}

\author{Cosimo Bambi}
\email[Corresponding author: ]{bambi@fudan.edu.cn}
\affiliation{Center for Field Theory and Particle Physics \& Department of Physics,
Fudan University, 220 Handan Road, 200433 Shanghai, China}

\swapnumbers
\begin{abstract}
In a previous paper, some of us studied general relativistic homogeneous 
gravitational collapses for dust and radiation, in which the density profile was
replaced by an effective density justified by some quantum gravity models.
It was found that the effective density introduces an effective pressure that
becomes negative and dominant in the strong-field regime. With this set-up, the 
central singularity is replaced by a bounce, after which the cloud starts expanding.
Motivated by the fact that in the classical case homogeneous and inhomogeneous
collapse models have different properties, here we extend our previous work to 
the inhomogeneous case. As in the quantum-inspired homogeneous collapse 
model, the classical central singularity is replaced by a bounce, but the 
inhomogeneities strongly affect the structure of the bounce curve and of the 
trapped region.
\end{abstract}

\pacs{04.20.Dw, 04.20.Jb, 04.70.Bw}
\keywords{Gravitational collapse, black holes, naked singularity}

\maketitle

\section{Introduction}

General relativistic gravitational collapses have been studied for many years since 
the pioneering work by Oppenheimer, Snyder and Datt~\cite{OSD} showed 
that a spherical matter cloud collapsing under its own weight leads to the formation 
of a black hole (BH). In this simple model, where the matter is described by 
homogeneous dust (i.e. pressureless) particles, the horizon forms at the boundary 
of the collapsing cloud before the formation of the central singularity. The system 
eventually settles to a Schwarzschild BH and the singularity remains inaccessible 
to far away observers. Since then, a lot of work has been done in order to understand 
the genericity and possible limitations of such a model. Singularity theorems by 
Hawking and Penrose~\cite{sing} show that under reasonable requirements for the 
matter content (i.e. energy conditions) if trapped surfaces do form then a singularity 
must form as well. Still they do not provide any information about how and when 
these singularities form. Further investigations showed that for certain matter profiles 
that satisfy standard conditions the singularity can form at the same time 
of the formation of the trapped surfaces and can thus be visible to far away observers
(see e.g.~\cite{review} and references therein for an overview of relativistic collapse).
The two most important features that arise from the study of the complete gravitational 
collapse of a massive cloud within the theory of general relativity are the trapped 
surfaces and the singularity.

It is usually thought that the appearance of spacetime singularities is a symptom
of the break down of classical general relativity, to be fixed by unknown quantum 
corrections. In Ref.~\cite{BMM}, such a possibility was explored and the homogeneous 
collapse of dust and radiation was re-analyzed in the light of corrections that might 
arise in the strong field regime, as obtained within some Loop Quantum Gravity (LQG) 
approaches~\cite{LQG-1,LQG-2,LQG-3}. 
The procedure is similar to the one followed in models of 
Loop Quantum Cosmology (LQC)~\cite{LQC}. The main result obtained in~\cite{BMM}
is that the singularity at the end of the collapse is removed and replaced by a bounce. 
The expanding phase that follows the collapsing phase after the bounce affects the 
structure of trapped surfaces in the sense that the event horizon of the Schwarzschild 
spacetime does not form, being replaced by an apparent horizon that exists for a 
finite time. These results appear in accordance with other studies carried out along 
the same line in several contexts (see for 
example~\cite{other1,other2,other3,other4,other5,other6}).

In the case of the gravitational collapse of an astrophysical object such as a star, 
the homogeneous dust model is highly unrealistic. Here we attempt to extend the 
analysis developed in Ref.~\cite{BMM} to the more realistic case of inhomogeneous dust. 
Since already in the fully classical case the structure of trapped surfaces and 
singularity is drastically altered by the introduction of inhomogeneities, it is worth 
investigating what happens in the quantum-inspired model. The presence of 
inhomogeneities in the classical case allows for the central region in which the 
singularity forms to be visible to far away observers. This suggests that the structure 
of the bounce and of the trapped surface will be also altered in the quantum-inspired
framework. We note that numerical studies of inhomogeneous gravitational collapse 
of scalar fields with LQG inspired corrections were reported in~\cite{new-r1,new-r2,new-r3}.

In cosmology, one may expect that the inhomogeneities arise from fluctuations at the 
quantum level of the gravitational field and the introduction of similar inhomogeneities 
in LQC models can be very difficult. Attempts to study inhomogeneous LQC models 
have been carried out by several authors~\cite{inhomogeneousLQC1,inhomogeneousLQC2,inhomogeneousLQC3,inhomogeneousLQC4}.
On the other hand, when we deal with the collapse of a massive object such as a star, 
we start with a matter distribution where inhomogeneities can be described at a purely 
classical level. Therefore we can consider an initial configuration given by a classical inhomogeneous dust ball that collapses under its own weight and consider the 
quantum-gravity effects at a semiclassical level only toward the end of the collapse.

The paper is organized as follows. In Section~\ref{classical}, we briefly review the 
formalism for the relativistic collapse of inhomogeneous dust matter. In Section~\ref{quantum}, 
we analyze how the relativistic picture is altered once quantum corrections in the 
strong field limit are considered. Finally, Section~\ref{conc} is devoted to a brief 
summary and discussion. In this paper, we use units in which $c=G_{\rm N}=1$ 
and absorb the factor $8\pi$ in the Einstein equations into the definition of the 
energy-momentum tensor.

\section{Classical collapse}\label{classical}

Here we assume that the collapse is spherically symmetric. Then the most general 
line element describing collapse in comoving coordinates can be written as
\be
ds^2=-e^{2\nu}dt^2+\frac{R'^2}{G}dr^2+R^2d\Omega^2 \; ,
\ee
where $d\Omega^2$ represents the two-dimensional metric on the unit two-sphere.
The metric functions $\nu(r,t)$, $G(r,t)$, and $R(r,t)$ are related to the physical density 
and pressures appearing in the energy-momentum tensor via the Einstein equations. 
The energy-momentum tensor in the comoving frame is diagonal and for a perfect 
fluid source depends only on density $\rho(r,t)$ and pressure $p(r,t)$. The Einstein 
equations can be written as
\bea \label{rho}
\rho&=&\frac{3M+rM'}{a^2(a+ra')}\; , \\ \label{p}
p&=&-\frac{\dot{M}}{a^2\dot{a}} \; ,\\ \label{nu}
\nu'&=&-\frac{p'}{\rho+p} \; , \\ \label{G}
\dot{G}&=&2\frac{\nu'r\dot{a}}{a+ra'}G \; ,
\eea
where we have absorbed the factor $8\pi$ into the definition of density and pressure. 
The scale factor $a(r,t)$ is a dimensionless quantity describing the rate of the collapse 
and is given by $R=ra$. The function $M(r,t)$ is related to the Misner-Sharp 
mass of the system $F=R(1-g_{\mu\nu}\nabla^\mu R\nabla^\nu R)$ (describing the amount of matter enclosed within the shell labelled by $r$ at the time $t$) via $F=r^3M$ 
and is given by
\be \label{misner}
M=a\left(\frac{1-G}{r^2}+e^{-2\nu}\dot{a}^2\right)\; .
\ee
Given the freedom to specify the initial scale, we choose the initial time $t_i=0$ such 
that $R(r,0)=r$, which implies $a(r,0)=1$. Matching with an exterior Schwarzschild 
or Vaidya spacetime is done at the comoving radius $r_b$ corresponding to the 
shrinking physical area-radius $R_b(t)=R(r_b,t)$~\cite{matching}.

The addition of an equation of state for the matter content that relates $p$ to $\rho$ 
provides the further relation to close the system of the Einstein equations. If no 
equation of state is provided, one is left with the freedom to specify one free function, 
still satisfying basic requirements of regularity and energy conditions. One
usually assumes that the matter content satisfies standard energy conditions (e.g. 
the weak energy conditions given by $\rho\geq 0$ and $\rho+p\geq 0$) and are 
regular and well behaved at the initial time at all radii. In this case, it is easy to 
prove that the singularity is reached for $a=0$ and it is a strong shell-focusing 
curvature singularity, where curvature invariants such as the Kretschmann scalar 
diverge. The curve $t_{ah}(r)$ that describes the apparent horizon is given by 
the condition $1-F/R=0$, which corresponds to $a(r,t_{ah}(r))=r^2M(r,t_{ah}(r))$, 
and it represents the time at which the shell labelled by $r$ becomes trapped.

\subsection{Homogeneous dust collapse}

The simplest possible model that one can obtain from the above set of equations 
is that of homogeneous pressureless matter. From the condition $p=0$, using 
Eq.~\eqref{p} we get $M=M(r)$. From the requirement that the density is homogeneous, 
namely $\rho=\rho(t)$, we get $M=M_0={\rm const.}$. Then Eq.~\eqref{nu} implies 
$\nu=\nu(t)$ and by a suitable reparametrization of the time we can set $\nu=0$. 
This leads to $\dot{G}=0$ in Eq.~\eqref{G} from which we get $G=f(r)$. The 
Misner-Sharp mass in Eq.~\eqref{misner} can be written as an equation of motion 
and we see that homogeneity implies that $f(r)=1+kr^2$, with $k={\rm const}$. 
The system is then fully solved once we integrate the equation of motion~\eqref{misner} 
written as
\be
\dot{a}=-\sqrt{\frac{M_0}{a}+k}\; .
\ee
In the simple case of marginally bound collapse (corresponding to particles having 
zero initial velocity at radial infinity) given by $k=0$, we obtain the solution 
for the scale factor
\be
a(t)=\left(1-\frac{3}{2}\sqrt{M_0}t\right)^{2/3}\; ,
\ee
where the integration has been performed with the initial condition $a(0)=1$. The 
singularity at the end of the collapse is simultaneous and occurs at the time
$t_s=2/3\sqrt{M_0}$, while the apparent horizon curve is given by $t_{ah}=t_s-2r^3M_0/3$. 
The horizon forms at the boundary of the cloud at the time 
$t_{ah}(r_b)<t_s$ and the singularity is therefore covered at any time.

\subsection{Role of inhomogeneities}\label{in-dust}

The introduction of perturbations in the classical density $\rho$ is equivalent to 
consider a mass profile $M$ that varies with $r$. Inhomogeneous models were 
first studied by Lema\`{i}tre, Tolman and Bondi~\cite{LTB}. From the Einstein 
equations, we obtain again $\nu=0$ and $G=f(r)=1+r^2b(r)$. The equation of 
motion becomes
\be
\dot{a}(r,t)=-\sqrt{\frac{M(r)}{a(r,t)}+b(r)}\; ,
\ee
and the scale factor for the marginally bound collapse case becomes
\be\label{a}
a(r,t)=\left(1-\frac{3}{2}\sqrt{M(r)}t\right)^{2/3}\; .
\ee
Now the singularity is not simultaneous any more. The time at which the 
shell labelled by $r$ becomes singular is given by the curve $t_s(r)=2/3\sqrt{M(r)}$, 
while the apparent horizon curve is given by $t_{ah}=t_s(r)-2r^3M(r)/3$. We now 
see that, depending on the behavior of the free function $M$, the structure of 
the singularity and of the apparent horizon curves can be very different. Given the 
continuity requirements that we must impose on $M$, it is reasonable to assume 
that close to the center the mass profile behaves like
\be\label{M(r)}
M(r)=M_0+M_2r^2+... \; .
\ee
To have a physically viable model that describes a realistic object, we 
would expect that the density is radially decreasing outward. This implies that 
the parameter $M_2$ in Eq.~\eqref{M(r)} is negative.

In the inhomogeneous case, the singularity forms at $r=0$ at the time $t_s(0)=2/3\sqrt{M_0}$ 
and outer shells become singular at later times. The behavior of the apparent horizon 
near the center is also determined by the value of $M_2$. For $M_2<0$, the 
apparent horizon forms at $r=0$ at the same time of the formation of the singularity 
and the outer shells become trapped afterward. In this case, it is easy to prove 
that the central singularity can be visible, at least locally, to far away observers 
(meaning that there exist families of null geodesics escaping from the singularity).
Also, given the nature of dust collapse (i.e. the absence of pressures), the boundary 
of the cloud can always be chosen at will thus making any locally naked singularity 
also globally naked~\cite{dust1,dust2,dust3,dust4,dust5}.

Whether such naked singularities can practically affect observations in a realistic 
scenario is an entirely different matter. In fact, toward the last stages of collapse,
if nothing happens to deviate from the classical relativistic picture, gravity dominates 
and densities are so high that for any practical purpose the radiation emitted from 
a collapsing object forming a naked singularity will be undistinguishable from that 
emitted from an object forming a BH~\cite{LingYao}. On the other hand, if quantum 
effects were to modify the picture of collapse close to the formation of the singularity, 
the fact that such a region of the spacetime is not trapped behind an horizon 
might bear important implications for the future development of the cloud.

\section{Quantum-inspired collapse} \label{quantum}

The introduction of inhomogeneities in the classical dust collapse drastically alters 
the structure of the singularity and of the trapped surfaces. It is thus reasonable to 
ask whether inhomogeneities will play an important role even in our quantum-inspired
model. There are different ways to introduce quantum corrections to the classical 
collapse in the strong field regime. Here we shall make use of a semiclassical treatment 
by assuming that the corrections to the Einstein tensor due to quantum effects
can be taken into account by replacing the matter source by an 
effective matter source. 
Therefore we will write the usual Einstein equations in the following form
\be
G_{\mu\nu}=T_{\mu\nu}^{\rm eff}\; ,
\ee
where $T_{\mu\nu}^{\rm eff} \rightarrow T_{\mu\nu}$ in the weak field 
limit and $T_{\mu\nu}$ is the classical energy-momentum tensor for dust. The specific 
form of $T_{\mu\nu}^{\rm eff}$ will depend on the specific approach to quantum gravity.
Of course $\nabla_\mu T^{\mu\nu}_{\rm eff} = 0$, but this is automatically 
satisfied in our approach because we will use the Einstein equations, which imply
the Bianchi identity, and we will not overconstrain the theory by imposing specific 
requirements for the matter content. We just demand that the standard framework is 
recovered in the weak field limit and we will check {\it a posteriori} if a reasonable 
interpretation for the matter content is possible in the strong field regime.
It is often believed that asymptotic freedom will play an important role at high densities 
in a way such that the gravitational interaction will diminish the density increases and 
infalling particles get closer. One way of modeling this behavior at a semiclassical level 
is to assume a variable coupling term $G_N$ (that in the classical scenario is Newton's 
constant), where $G_N$ will depend on $\rho$.

A similar approach is used to construct bouncing cosmological models within LQG. 
A homogeneous Friedmann-Robertson-Walker model is altered in such a way that 
the big bang singularity is replaced by a bounce~\cite{LQC}. In cosmological models, 
one expects that the large scale structures form from small inhomogeneities that are
originated in the early Universe at a quantum level. Nevertheless, introducing 
inhomogeneities at a quantum level is not an easy task and there are difficulties due 
to the fact that we do not posses a viable theory of quantum gravity yet. On the other 
hand, for a collapse scenario, as already mentioned, the initial state of the system can 
be considered as purely classical and all the quantum corrections can be neglected at the 
initial time. The inhomogeneities that we consider are macroscopic perturbations in the
matter distribution and appear in the stress energy tensor, where the classical $\rho$ 
depends on $r$. We then follow the evolution of a classical inhomogeneous dust 
collapse to the point where quantum corrections become important and we treat these 
corrections at a semiclassical level modifying the stress energy tensor ``shell by shell''.
Following Ref.~\cite{BMM}, we assume that the effective density can be written in the 
form
\be\label{lqg}
\rho^{\rm eff}=\rho\left(1-\frac{\rho}{\rho_{\rm cr}}\right)\; .
\ee
Here $\rho_{\rm cr}$ plays the role of a critical density associated with the minimum 
scale of collapse and can be related to the limit in which the gravitational attraction 
vanishes. The presence of the correction term in the effective energy density will induce 
an effective pressure in the dust collapse scenario that will become negative as the collapse approaches the critical stage. This effective pressure describes how the system approaches asymptotic safety. In the same manner, the mass function $M(r)$, which is related to the 
total Schwarzschild mass measured by far away observers 
$\mathrm{M}_{\rm Sch}=r_b^3M(r_b)/2$, is replaced by a variable effective mass 
$M^{\rm eff}(r,t)$ that decreases as the collapse progresses. Then following the standard 
matching conditions for classical general relativity one can perform the matching at the 
boundary with a radiating Vaidya exterior, which again has to be understood in the 
effective picture.

\subsection{Homogeneous case}

\begin{figure}[hhh]
\begin{center}
\includegraphics[scale=0.30]{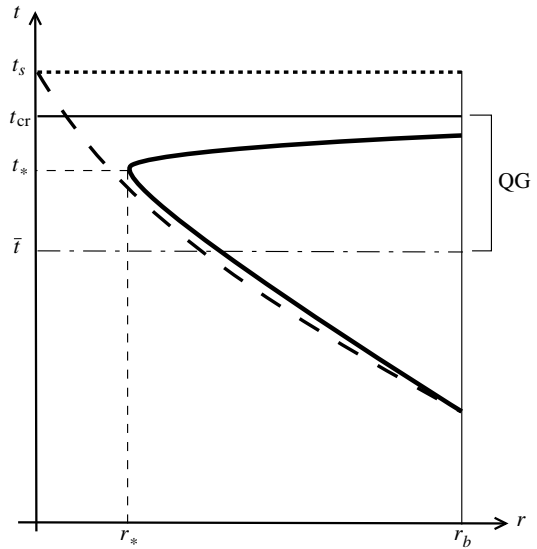}
\end{center}
\vspace{-0.4cm}
\caption{Schematic illustration of the homogenous bounce in comoving coordinates. 
The collapse follows the classical model until the time $\bar{t}$ at which the quantum gravity regime becomes important. The semiclassical apparent horizon (continuous thick line) 
separates from the classical one (dashed thick line), reaches a minimum $r_\star$ 
at the time $t_\star$ and then diverges. All the shells bounce at the same time 
$t_{\rm cr}$ before the classical time of the singularity $t_s$.}
\label{fig1}
\end{figure}

A model for homogeneous dust collapse inspired by the LQG corrections was investigated 
in Ref.~\cite{BMM} (see also Fig.~\ref{fig1}). With the initial condition $a(0)=1$, one finds 
the following solution for the scale factor
\be
a(t)=\left[a_{\rm cr}^3+\left(\sqrt{1-a_{\rm cr}^3}-
\frac{3\sqrt{M_0}}{2}t\right)^2\right]^{1/3} \; ,
\ee
where we have defined $a_{\rm cr}^3=3M_0/\rho_{\rm cr}$. It is easy to see that as the 
critical density goes to infinity we retrieve the classical homogeneous dust collapse model.

For the homogeneous semiclassical model, all the shells bounce at the same comoving 
time $t_{\rm cr}=2\sqrt{1-a_{\rm cr}^3}/(3\sqrt{M_0})$. Therefore, as a consequence of 
the homogeneity, we have a simultaneous bounce replacing the simultaneous singularity. 
The apparent horizon is again defined as the curve $t_{ah}(r)$ for which 
$a(r,t_{ah}(r))=r^2M^{\rm eff}(r,t_{ah}(r))$. In the homogeneous case, one can see that 
the apparent horizon initially behaves like in the classical case, reaches a minimal radius 
$r_\star$ at the time $t_\star=t_{\rm cr}(1-\sqrt{3a_{\rm cr}^3}/\sqrt{1-a_{\rm cr}^3})$ and 
then re-expands crossing the boundary again before the time of the bounce. At the time 
of the bounce, we reach the asymptotic freedom regime in which the gravitational 
force vanishes. After $t_{\rm cr}$, the cloud re-expands following a dynamics that is 
symmetrical to the collapsing case. Another trapped region forms in the expanding phase due to the fact that the gravitational attraction grows as the system leaves the asymptotic safe regime and eventually the whole cloud disperses to infinity.

\subsection{Inhomogeneous case}

An exact procedure to deal with inhomogeneities at the level of quantum gravity is 
presently not known. Luckily, for the purpose of studying a gravitational collapse 
we can consider a cloud that is already inhomogeneous in the weak field and thus 
begin with classical inhomogeneities as described in Section~\ref{in-dust}. The 
only guidelines we keep in mind when we introduce inhomogeneities are that 
we want to recover the classical case when the critical density goes to infinity and 
we want to recover the homogeneous case when the density perturbations go to 
zero. Nevertheless, even with these great simplifications, treating the problem 
analytically can prove to be too difficult. In what follows, we shall thus restrict our 
attention to the vicinity of the center of the cloud, by performing a Taylor expansion
 of all the relevant quantities near $r=0$. This is possible due to the regularity of 
 the functions involved even close to the classical singularity. 
 We stress that in this way we are not assuming the existence of a 
 bounce replacing the classical singularity. Indeed the same approach is used to 
 study the formation of singularities in the classical case, where one can describe 
 the collapse up to the critical time $t_{\rm cr}$~\cite{review}.
 Here we use the same strategy and 
 eventually we found a bounce: thanks to the regularity of the solution, {\it a posteriori}
 we can say that the model holds even after the bounce. By 
 expanding all the functions in the vicinity of $r=0$, we are able to reduce the system 
 of five coupled partial differential equations given by Eqs.~\eqref{rho}-\eqref{misner} 
 to a system of two coupled ordinary differential equations. Using Eq.~\eqref{rho} for 
 the definition of the effective density in Eq.~\eqref{lqg}, we obtain the effective mass 
 function $M^{\rm eff}(r,t)$ that can be expanded in powers of $r$ as
\be
M^{\rm eff}(r,t)=M^{\rm eff}_0(t)+M^{\rm eff}_2(t)r^2 +... \; ,
\ee
with
\bea \label{Meff0-1}
M^{\rm eff}_0&=& M_0\left(1-\frac{K}{a_0^3}\right) \; , \\ \label{Meff2-1}
M^{\rm eff}_2&=&M_2\left(1-2\frac{K}{a_0^3}\right)+3M_0\frac{Ka_2}{a_0^4}\; ,
\eea
where we have defined $K=3M_0/\rho_{\rm cr}$ and expanded the scale factor as
\be
a(r,t)=a_0(t)+a_2(t)r^2+...\; .
\ee

In order to write the equation of motion for the scale factor up to the second order, we 
need now to solve the full system of the Einstein equations in the effective picture. 
The dependence on $t$ of the effective mass function will induce the presence of a 
non-vanishing effective pressure that can be expanded as 
$p^{\rm eff}=p^{\rm eff}_0+p^{\rm eff}_2r^2+...$, where
\bea
p^{\rm eff}_0&=& -\frac{3M_0K}{a_0^6}  \; , \\
p^{\rm eff}_2&=& -\frac{6M_2K}{a_0^6}+\frac{18M_0Ka_2}{a_0^7}\; .
\eea
From the remaining Eqs.~\eqref{nu} and \eqref{G} we get
\bea
\nu&=&\nu_2r^2+...=-\frac{p^{\rm eff}_2}{\rho^{\rm eff}_0+p^{\rm eff}_0}r^2+... \; , \\
G&=&b(r)e^{2A}\; ,
\eea
with $A$ defined by
\be
\dot{A}:=\nu'\frac{r\dot{a}}{a+ra'}=\dot{A_2}r^2+... \; .
\ee
If we restrict ourselves to the marginally bound case given by $b=1$, we can 
expand $G$ as $G=1+2A_2r^2+...$ and we obtain
\bea
\nu_2&=&\frac{2K}{a_0^3} \frac{\frac{M_2}{M_0}-\frac{3a_2}{a_0}}{1-\frac{2K}{a_0^3}}  \; , \\
A_2&=&2\int_0^t\nu_2\frac{\dot{a}_0}{a_0}d\bar{t}\; .
\eea

\begin{figure}
\begin{center}
\includegraphics[scale=0.90]{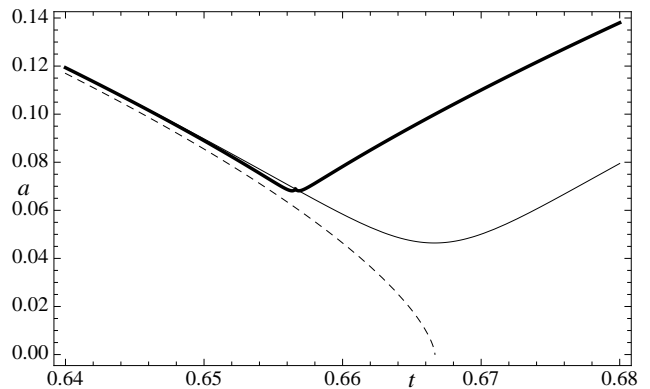}
\end{center}
\caption{The scale factor in the classical case (dashed line), in the homogeneous 
semiclassical case (thin line) and in the inhomogeneous semiclassical case for a fixed 
small value of $r$ (thick line, $r=0.01$). The following numerical values have been chosen: $M_0=1$, 
$M_2=-0.1$, and $3M_0/\rho_{\rm cr}=0.0001$.}
\label{fig2}
\end{figure}

Assuming that higher order terms are negligible, we finally get the expansion of the 
equation of motion~\eqref{misner} written order by order in the effective picture as
\bea \label{Meff0-2}
M^{\rm eff}_0&=&a_0(-2A_2+\dot{a}_0^2)\; ,  \\ \label{Meff2-2}
M^{\rm eff}_2&=&
a_2(-2A_2+\dot{a}_0^2)+2a_0\left[\dot{a}_0\dot{a}_2-\nu_2\dot{a}_0^2\right]\; .
\eea
In the limit for $K=0$ (corresponding to $\rho_{\rm cr}$ going to infinity), we retrieve 
the classical inhomogeneous collapse model, while in the limit for $M_2=0$ we obtain 
the homogeneous quantum-inspired model discussed in Ref.~\cite{BMM}. When we 
combine the above equations with Eqs.~\eqref{Meff0-1} and \eqref{Meff2-1}, we get 
the two equations of motion that need to be solved in order to obtain the expansion of 
the scale factor in the inhomogeneous quantum-inspired model. From the first one 
we get
\be\label{a0}
\dot{a}_0^2=\frac{M_0}{a_0}\left(1-\frac{K}{a_0^3}\right)+2A_2 \; ,
\ee
which, after we derive again with respect to $t$ and substitute for $\dot{A}_2$ gives
\be
\ddot{a}_0=-\frac{M_0}{2a_0^2}+\frac{2M_0K}{a_0^5}
+\frac{4K}{a_0^4-2Ka_0}\left(\frac{M_2}{M_0^2}-\frac{3a_2}{a_0}\right) \; .
\ee
Then the second one leads to
\be\label{a2}
\dot{a}_2=\frac{M_2}{2a_0\dot{a}_0}\left(1-\frac{2K}{a_0^3}\right)
-\frac{M_0a_2}{2a_0^2\dot{a}_0}\left(1-\frac{4K}{a_0^3}\right)+\nu_2\dot{a}_0 \; .
\ee
Notice that in the inhomogeneous case the scale factor at zero order given by $a_0$, 
as the solution of Eq.~\eqref{a0}, is different from $a$ in the homogeneous case. This is 
due to the non linearity of the Einstein equations that adds the term $2A_2$ in 
Eq.~\eqref{a0}, which vanishes in the homogeneous limit. This is reflected in a different 
time of the bounce for the central shell with respect to $t_{\rm cr}$ in the homogeneous 
case, provided that the system is normalized with the same scaling at the initial time 
(see Fig.~\ref{fig2}).

Our analysis is valid in the limit  of small $r$, for which we can assume that all the
higher order terms are negligible. In the general case, $M_0$ sets the scale for 
the collapse scenario, and this approximation breaks down at a certain radius for 
any given value of $M_2$ and $\rho_{\rm cr}$. Classically, the limit of validity of 
the small $r$ approximation is determined by $M_2$ only. Another important
issue concerns the possibility of shell crossing singularities. The latter are weak 
curvature singularities that arise when different collapsing shells overlap~\cite{shellcross}.
They are obtained from the curvature scalars when the condition $a+ra'=0$ is 
satisfied, but they do not signal geodesic incompleteness of the spacetime, which 
can be indeed extended through them. Shell crossing singularities do not appear 
in the case of the classical dust collapse if the energy density profile is homogeneous 
or radially decreasing outward. Nevertheless, for other density profiles and 
whenever pressures are present in the cloud one needs to check that no shell 
crossing singularities occur during the collapse. In the quantum-inspired scenario, 
in general the situation is made even more complicated by the fact that reflected 
shells will lead to caustics when overlapping with infalling shells, and these will 
also be indicated by shell crossing singularities. However, in the model studied 
here the bounce occurs first at the outer shells and thus if shell crossing 
singularities do happen they are confined outside the regime of validity of our
``small $r$'' approximation.

One final point to mention concerns the classical physical density. We have 
considered here a classical density given by an expansion where $\rho$ satisfies 
the Einstein field equations. In general, it is possible that the classical relativistic 
expression for $\rho$ will not hold as the density approaches the critical value. 
This, in turn, will affect the form of the effective density derived in the semiclassical 
scenario. Since one does not know in principle how to write the modified density, 
and since we know that for $\rho_{\rm cr}$ going to infinity we must recover the 
classical case satisfying classical field equations, it makes sense to consider
$\rho=\rho_{\rm GR}+\epsilon(t)$, where $\rho_{\rm GR}$ is the relativistic energy 
density given by Eq.~\eqref{rho} and $\epsilon(t)$ is an arbitrary function that 
depends on $\rho_{\rm cr}$ and accounts for such modifications. The form of 
$\epsilon$ will then depend on the specific approach to the inhomogeneous system in 
quantum gravity. Nevertheless, $\epsilon$ must be negligible in the weak field 
regime (i.e. close to the initial time) and must go to zero as the critical density goes 
to infinity. It seems therefore reasonable to assume at first instance that the effect 
of $\epsilon(t)$ is negligible at any time. For this reason, and for simplicity, in 
order to minimize the number of free parameters in the analysis we have chosen 
to take $\epsilon=0$ during the whole dynamical evolution. Solving the coupled 
system of equations given by \eqref{a0} and \eqref{a2} analytically could prove to 
be impossible but we can still understand the behavior of collapse near the center 
by solving the system of equations numerically.

\subsection{Bounce and trapped surfaces}\label{bounce}

\begin{figure}
\begin{center}
\includegraphics[scale=0.90]{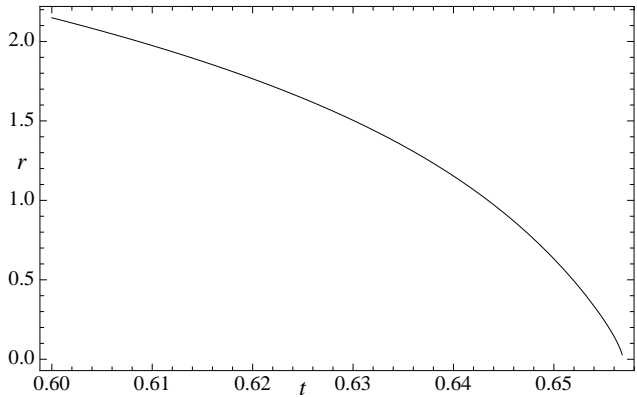}
\end{center}
\caption{The bounce curve $r_{\rm cr}(t)$ in the inhomogeneous case. 
The following numerical values have been chosen: $M_0=1$, $M_2=-0.1$, and
$3M_0/\rho_{\rm cr}=0.0001$.}
\label{fig3}
\end{figure}

Because of the presence of inhomogeneities, the behavior of the collapsing cloud 
is affected ``shell by shell'', and it is easy to see that the time of the bounce is
different for every shell. We can define the ``bounce curve'' $t_{\rm cr}(r)$ from 
the bounce condition
\be
\dot{a}(r,t_{\rm cr}(r))=0 \; .
\ee
The crucial element that distinguishes the bounce from the homogeneous case is 
that $t_{\rm cr}(r)$ (or inversely $r_{\rm cr}(t)$) is not a constant (see Fig.~\ref{fig3}).

This means that the asymptotic freedom regime 
is achieved at different times for each shell and thus the gravitational attraction does 
not vanish entirely at a specific time. An important consequence of the presence 
of the bounce curve is that the outer shells (still considering only shells close to 
the center) bounce before the inner shells, as opposed to the classical scenario 
where the singularity forms initially at the center when $M_2<0$. Then shell crossing 
singularities are not present near the center of the cloud. Nevertheless, there will 
be a certain radius at which the approximation for small $r$ ceases to be valid. 
Expanding shells coming from the bounce will intersect the outer shells that still 
follow classical collapse causing caustics, shell crossing singularities and a 
breakdown of the model.

\begin{figure}[ttt]
\begin{center}
\includegraphics[scale=0.90]{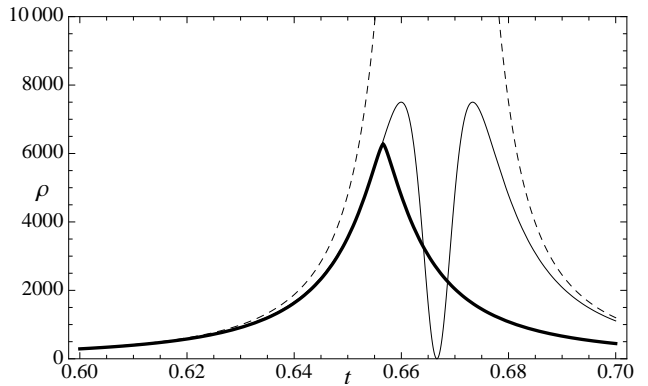}
\end{center}
\caption{Energy density in the classical scenario (dashed line), in the homogeneous 
semiclassical case (thin line) and in the inhomogeneous semiclassical case for fixed small value of $r$ (thick line, $r=0.01$). The following numerical values have been chosen: $M_0=1$, $M_2=-0.1$, and $3M_0/\rho_{\rm cr}=0.0001$.}
\label{fig4}
\end{figure}

The fact that $t_{\rm cr}$ is not constant implies also that the minimum value for 
the scale factor is different for every collapsing shell, and we thus have $a_{\rm cr}(r)$ 
with the smallest value obtained for the central shell. A consequence of this fact is 
that the effective density does not vanish everywhere at a specific time, as opposed to the homogeneous case in which $\rho^{\rm eff}=0$ at the time of the bounce (see Fig.~\ref{fig4}). Nevertheless, 
$\rho^{\rm eff}$ decreases as we approach the bounce and the effect of $\rho_{\rm cr}$ 
becomes more important than the effect of the inhomogeneity $M_2$ in this limit. 
This is reflected in the equations in the fact that the profile for the energy density 
goes from being decreasing radially close to the initial time to being increasing 
radially close to the time of the bounce.

\begin{figure}[hhh]
\begin{center}
\includegraphics[scale=0.30]{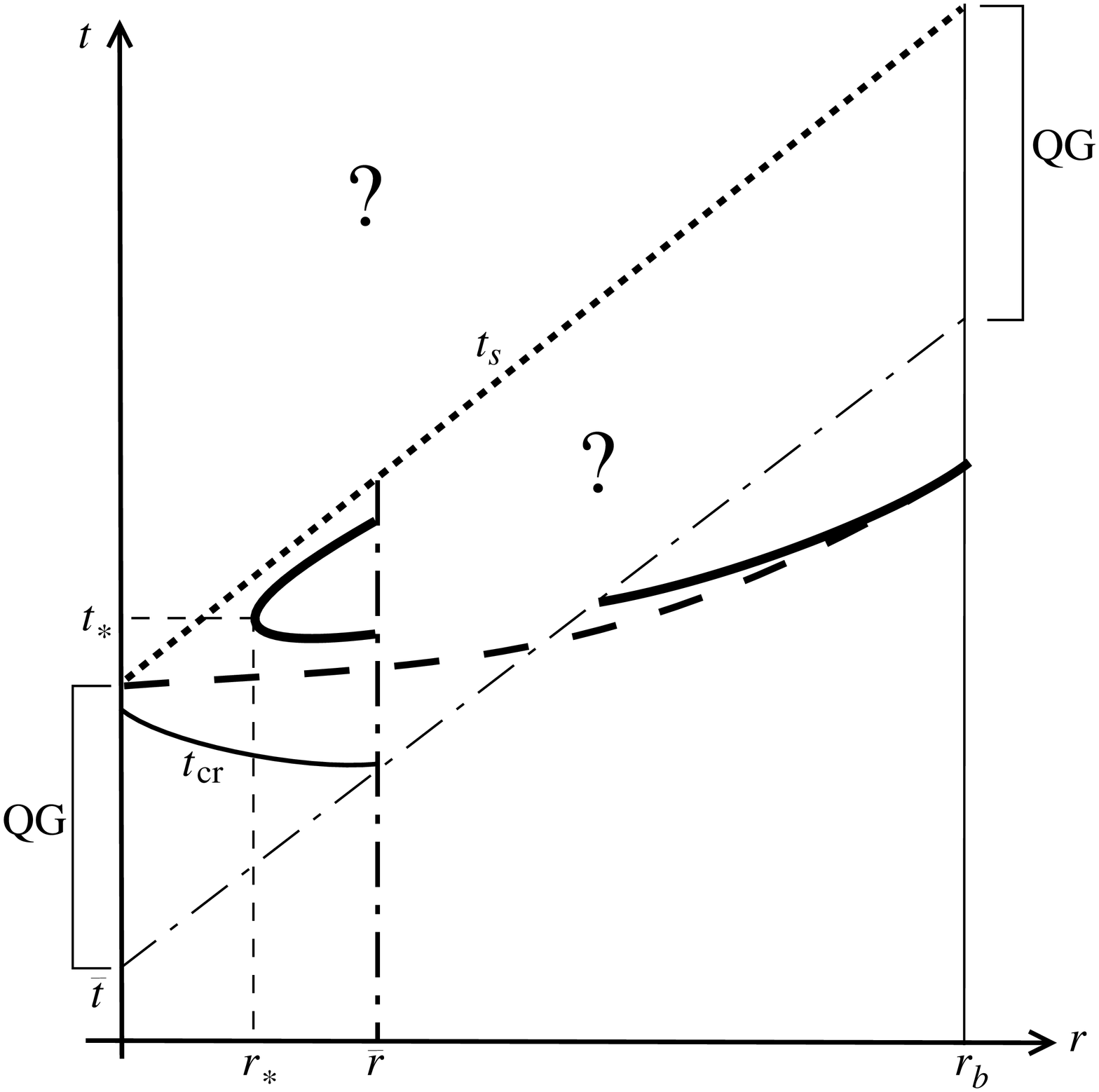}
\end{center}
\vspace{-0.4cm}
\caption{Schematic illustration of the inhomogenous bounce in the comoving frame. Collapse follows the classical model in the weak field. The strong field region is achieved at the center at an earlier time with respect to the boundary. The classical singularity curve $t_s(r)$ does not occur.
The semiclassical apparent horizon (continuous thick line) is close to the classical apparent horizon (dashed thick line) near the boundary in the weak gravity regime. Near the center the semiclassical apparent horizon reaches a minimum $r_\star$ 
at the time $t_\star$ and then re-expands. Shells bounce at different times and the bounce curve $t_{\rm cr}(r)$ (continuous thin line) is decreasing near $r=0$. We do not know the behavior of the bounce curve in the quantum-gravity region for large $r$ (inside the dotted-dashed line for $r>\bar{r}$) and we do not know the behavior of the cloud at late times after the bounce when the expanding shells meet the collapsing ones.}
\label{fig5}
\end{figure}

Similarly to the classical inhomogeneous collapse model, the structure of formation of the 
trapped surfaces is given by the curve $t_{ah}(r)$ that represents the time at which the 
shell labelled by $r$ becomes trapped. This is given implicitly by
\be \label{app-hor}
a(r,t_{ah}(r))=r^2M^{\rm eff}(r,t_{ah}(r))\; .
\ee
In the classical inhomogeneous case with $M_2<0$, the horizon forms initially at the 
time of formation of the singularity and then ``propagates'' outward meeting the event 
horizon at the boundary at a later time. Once we consider the semiclassical picture, 
the singularity is replaced by a bounce, the action of gravity is diminished approaching 
asymptotic freedom and the formation of trapped surfaces is delayed. The inverse of $t_{ah}(r)$ can be obtained from Eq.~\eqref{app-hor} by solving the quadratic equation
\be
r^4M^{\rm eff}_2+r^2(M^{\rm eff}_0-a_2)-a_0=0 \; .
\ee
It is easy to see that since $a_0$ has a minimum the apparent horizon will not pass through the shell $r=0$ at any time. Therefore, like in the homogeneous case, we see that the apparent horizon behaves like the classical one in the weak field regime while it reaches a minimum value $r_\star$ at the comoving time $t_\star$ given by $\dot{r}_{ah}(t_\star)=0$. Then by numerically evaluating the time $\tilde{t}$ at which the central shell bounces we can see that $t_\star>\tilde{t}$ and therefore, close to the center, the bounce curve is not trapped inside the horizon (see Fig.~\ref{fig5}).

\section{Concluding remarks}\label{conc}

Strictly speaking, a BH is defined as a spacetime region causally disconnected 
to future null infinity and the event horizon is the boundary of such a region. In the present paper, 
we have shown that, contrary to the classical relativistic picture in which the collapse 
leads inevitably to the formation of a BH, such a final outcome can be prevented 
when one deals with the semiclassical corrections that are thought to arise in the 
strong field regime. The singularity at the end of the collapse does not form and 
instead the collapsing cloud bounces and enters a phase of re-expansion. Let us
notice that the non-formation of an event horizon, and therefore of a BH, can be 
understood in the effective framework as arising from the fact the exterior spacetime
is not described by the Schwarzschild solution, but by a Vaidya-like spacetime,
with an effective infalling flux of negative energy (see e.g. the discussion 
in~\cite{other1}).

When we consider a homogeneous dust cloud, the bounce occurs simultaneously 
(in the comoving frame), thus realizing an instant of complete asymptotic freedom 
where the gravitational force vanishes. In the homogeneous picture, a trapped 
region forms before the bounce and even though the timescale of the process 
for the comoving observers is short, of order of the dynamical scale of the 
system $M_{\rm Sch}$, for observers at spatial infinity the process appears much 
slower and the object can ``mimic'' a classical BH for a long time.

Introducing inhomogeneities at a semiclassical level drastically alters the scenario in 
the strong field regime (much like inhomogeneities can alter the structure of trapped 
surfaces in classical collapse). The bounce is not simultaneous anymore and the 
cloud never reaches a stage of complete asymptotic freedom, since different shells 
reach the critical density at different times. Interestingly, close to the center of the 
cloud, the outer shells bounce before the inner ones and the bounce is not 
accompanied by the formation of any trapped surface. Any classical apparent 
horizon that might form near the boundary of the cloud in the weak field regime is 
then bound to be swept away by the expanding inner shells after the bounce. As a 
consequence the high density region in which quantum gravitational effects become 
important is not covered by the horizon in this case. The possible observational 
consequences of the existence of exotic compact objects in the Universe have been a matter of 
great interest in recent years (see for 
example~\cite{obs1,obs2,obs3,obs4,obs5,obs6,obs7}). 
The present work suggests 
that quantum effects can alter the classical BH formation paradigm thus leaving 
open the possibility for the existence of a window on new physics in astrophysical phenomena.


\begin{acknowledgments}
This work was supported by the NSFC grant No.~11305038, 
the Shanghai Municipal Education Commission grant for Innovative 
Programs No.~14ZZ001, the Thousand Young Talents Program, 
and Fudan University.
\end{acknowledgments}


\end{document}